 \documentclass[final,1p,times]{elsarticle}
\usepackage{hyperref}

\usepackage{fancyhdr}
\usepackage{graphicx}
\graphicspath{{./}{./images/}{./images/AutresFormats/}}
\usepackage{epstopdf}
\usepackage[lofdepth,lotdepth]{subfig}

\usepackage{amssymb}
\usepackage{amsthm}
\usepackage{amsmath}
\usepackage{amsfonts}

\usepackage{paralist}

\journal{Applied Acoustics}

\begin{document}

\begin{frontmatter}



\title{Design and implementation of a multi-octave-band audio camera for realtime diagnosis}


\author{C.~Vanwynsberghe}
\ead{charles.vanwynsberghe@upmc.fr}

\author{R.~Marchiano}
\ead{regis.marchiano@upmc.fr}

\author{F.~Ollivier}
\ead{francois.ollivier@upmc.fr}

\author{P.~Challande}
\ead{pascal.challande@upmc.fr}

\author{H.~Moingeon}

\author{J.~Marchal}
\ead{jacques.marchal@upmc.fr}

\address{Sorbonne Universit\'{e}s, UPMC Univ Paris 06, CNRS, UMR 7190 Institut Jean Le
Rond d'Alembert, F-78210 Saint Cyr l'\'{E}cole, France.}

\begin{abstract}
Noise pollution investigation takes advantage of two common methods of diagnosis: measurement using a Sound Level Meter and acoustical imaging. The former enables a detailed analysis of the surrounding noise spectrum whereas the latter is rather used for source localization. Both approaches complete each other, and merging them into a unique system, working in realtime, would offer new possibilities of dynamic diagnosis. This paper describes the design of a complete system for this purpose: imaging in realtime the acoustic field at different octave bands, with a convenient device. The acoustic field is sampled in time and space using an array of MEMS microphones. This recent technology enables a compact and fully digital design of the system. However, performing realtime imaging with resource-intensive algorithm on a large amount of measured data confronts with a technical challenge. This is overcome by executing the whole process on a Graphic Processing Unit, which has recently become an attractive device for parallel computing.
\end{abstract}

\begin{keyword}
Acoustic imaging \sep realtime \sep GPU \sep Microphone array\sep MEMS

\end{keyword}

\end{frontmatter}
\thispagestyle{fancy} 

\section{Introduction}
Acoustic studies in various domains such as car, aircraft or train design, are mainly concerned by noise reduction. The latter is becoming stricter by both respect of regulations and passenger comfort. Most of these regulations go by standardized experiment protocols, and need the use of a Sound Level Meter (SLM). This instrument is the standard acoustic device for diagnosis, and gives an accurate description of the surrounding noise in terms of acoustic power. It provides the overall Sound Pressure Level (SPL), and octave or third-octave band levels for a finer spectral description. On another hand, source localization becomes an important tool for developers who want to reduce noise pollution from the physical origin. But no standardized protocol exists for such methods though some studies tend to it \cite{Kook2000}. Anyway acoustical imaging remains important, but has been limited by both hardware complexity and a strong need for computing resources. 

However, two recent technologies open up new perspectives for imaging systems. First, digital MicroElectroMechanical-Systems (MEMS) microphones are initially meant for general use devices implying speech acquisition \textit{e.g.} telephones. But their high convenience for electronic design enables more versatile applications such as building an acoustic imaging array \citep{Hafizovic2012}. Indeed with these full digital components, the global hardware system is simplified. Secondly, General Purpose Graphic Processing Units (GPU) become an excellent solution for high parallel computing while being cost-effective and compact. The standard beamforming (BF) imaging algorithm is well adapted to parallel architecture and gives access to a fast enough code execution for a realtime visualization \cite{Nilsen2009}. This has already been proved by several studies in ultrasound imaging \cite{Martin-Arguedas2012,Chen2012a,Asen2014,Asen2012,Chen2011} and in underwater acoustic imaging \cite{Buskenes2014, Buskenes2013}.

Taking advantage of these recent technologies, this paper describes a complete imaging system, from the data acquisition hardware to the signal processing and acoustic images display. Then this diagnosis tool intends to give an analysis consistent with classical SLM data. It maps, in realtime, acoustic levels according to standard octave bands. A previous study has explored the multi-frequency band imaging in realtime using GPU for versatile applications \cite{ODonovan2007}, using a classic acquisition system architecture.

The present work goes through two steps. First the hardware system is described; some calibration experiments performed to validate the suitability of MEMS microphones for the application in sight are also presented. Secondly, the design of the array beamformer is described, as well as its implementation on GPU. The achievability of realtime multiband imaging is discussed. Finally, we perform experimental tests in two different scenarios, to confirm the relevance of the presented diagnosis system.


\section{The data acquisition system}
\subsection{Hardware architecture}
\begin{figure}
\centering
\includegraphics[width=0.4 \paperwidth]{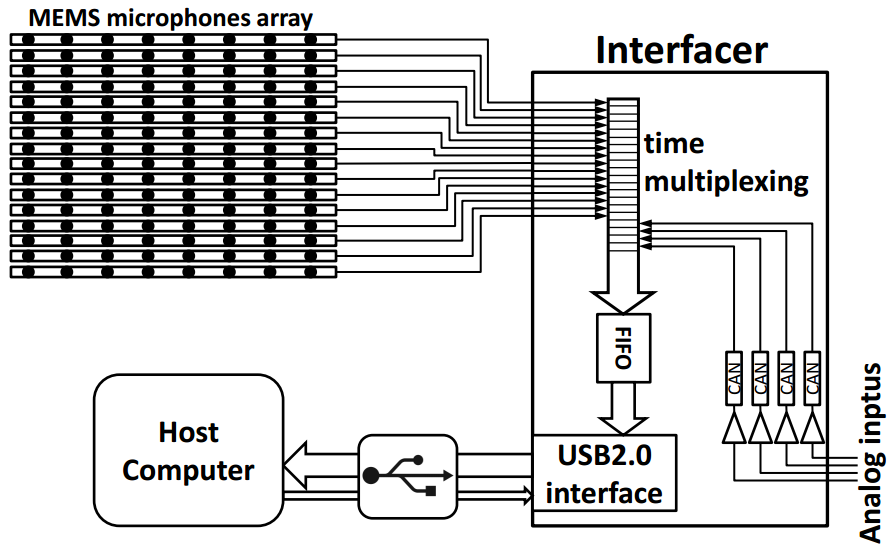}
\caption{Hardware architecture of the data acquisition system}
\label{fig:diagram_system}
\end{figure}

The acquisition system consists of two parts: the microphones array and an interface for communication with a host computer, as shown in figure \ref{fig:diagram_system}. The array is made of 16 sets of 8 MEMS microphones, resulting in a 128 elements antenna. Its geometry is arbitrarily user configurable. The component chosen for pressure measurement is the ADMP441 microphone developed by Analog Device \cite{admp441}. It allows a fully digital design of the system. Indeed the microchip includes the whole instrumentation chain, \textit{ie} the transducer, an amplifier and a 24 bits $ \Sigma \Delta$ converter. Compared with current systems, such an integrated sensor leads to a more condensed and simplified architecture.

The whole 128-microphone set is linked to the interface using the I2S serial protocol. It is driven with a user configurable sampling rate ($f_s$) common to the 128 acoustic channels, and synchronously sends the digital data. However the  data flow must fit to standard computers for both convenience and compatibility issues. Therefore, the 24-bit integer samples are converted to 32-bit floats before multiplexing  and transfer via the USB 2.0 serial bus. Finally a single time sample from the array consists of 512 Bytes, resulting in a total data flow rate of $ 512 f_s$~Byte/s from the interface to the computer. With a sampling frequency $f_s =50$~kHz the USB 2.0 bus is able to achieve a $25$~MByte/s rate in \textit{full-speed} mode.

Finally, the host computer receives the data to be processed for realtime imaging, and optionally saves them for further post-processing.

\subsection{Acoustic assessment of the MEMS microphones}
The ADMP441 microphones are initially intended for voice acquisition applications of general use \cite{admp441}, however their performances for aerial acoustic imaging applications are unknown. Previous works have made use of such components \cite{Hafizovic2012} but without concern for imaging. In this section the performances of these general purpose microphones are established in the specific field of acoustic imaging. Since the component datasheet does not provide accurate values for these particular specifications, a set of calibration experiments is performed.

\subsubsection{Design of experiments}

Four important specifications are investigated:
\begin{inparaenum}[i)]
\item sensitivity,
\item directivity,
\item frequency response and
\item self noise.
\end{inparaenum}
Indeed the imaging algorithm rely on strong hypothesis: all the sensors are supposed to have the same sensitivity and to be omnidirectional; meeting at best these criteria guarantees better reconstruction capability. Besides, frequency response determines the spectral limits of the diagnosable sound field, and self noise sets the hearing threshold of the system.

\begin{figure}
\centering
\includegraphics[width=0.3 \paperwidth]{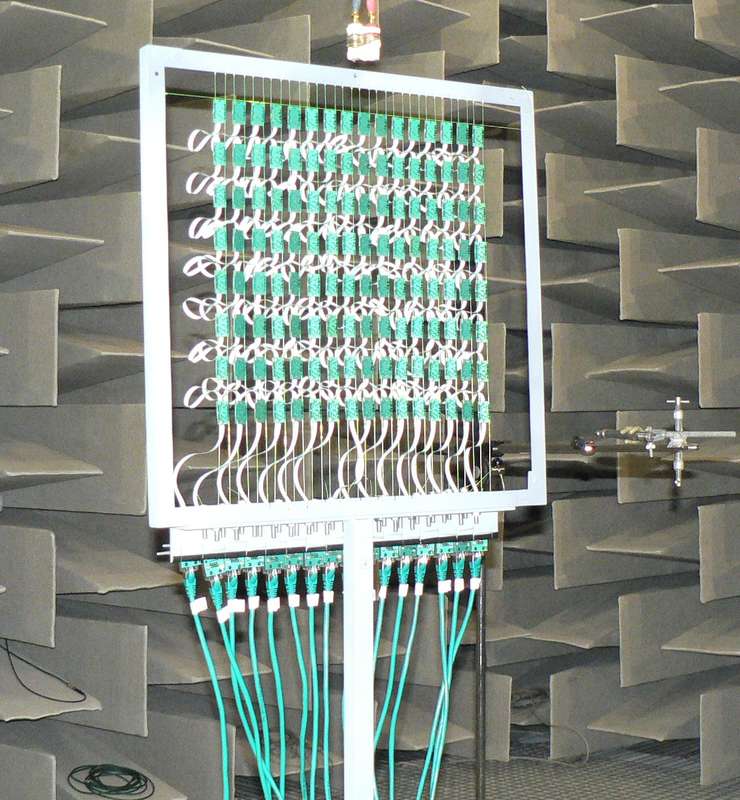}
\caption{MEMS assessment experimental setup in the anechoic chamber LNE - Frame dimensions: 50~cm~x~50~cm}
\label{fig:LNE}
\end{figure}

The experiments are performed in the anechoic chamber of the \textit{Laboratoire National d'Essai} (LNE) \cite{LNE}. The MEMS microphones are assembled on a compact frame (fig. \ref{fig:LNE}). This frame is located $5$~m away from a controlled source. This configuration allows to consider the impinging wave front to be of constant amplitude over the array, which makes a statistical analysis possible.

Using a standard loudspeaker emitting a white noise signal, 128 SPLs are measured, directly revealing the sensitivity variation of the microphones. Using the same source and rotating the frame over 180 degrees with a 10 degrees step, 128 azimuthal directivity patterns are derived.
The frequency response is referenced to a reference microphone type \textit{Bruel \& Kjaer 4190}. The transfer function estimation method is followed \cite{bendat1980}, using an exponential chirp to provide high signal to noise ratio (SNR). Finally self noise is evaluated by acquiring component outputs in silence.

\subsubsection{Results and discussion}

\begin{figure}
\begin{center}
	\subfloat[Histogram of 128 measured SPL referred to $2\cdot10^{-5}$ Pa, gaussian fitting in red]{
	\includegraphics[width=0.4\textwidth]{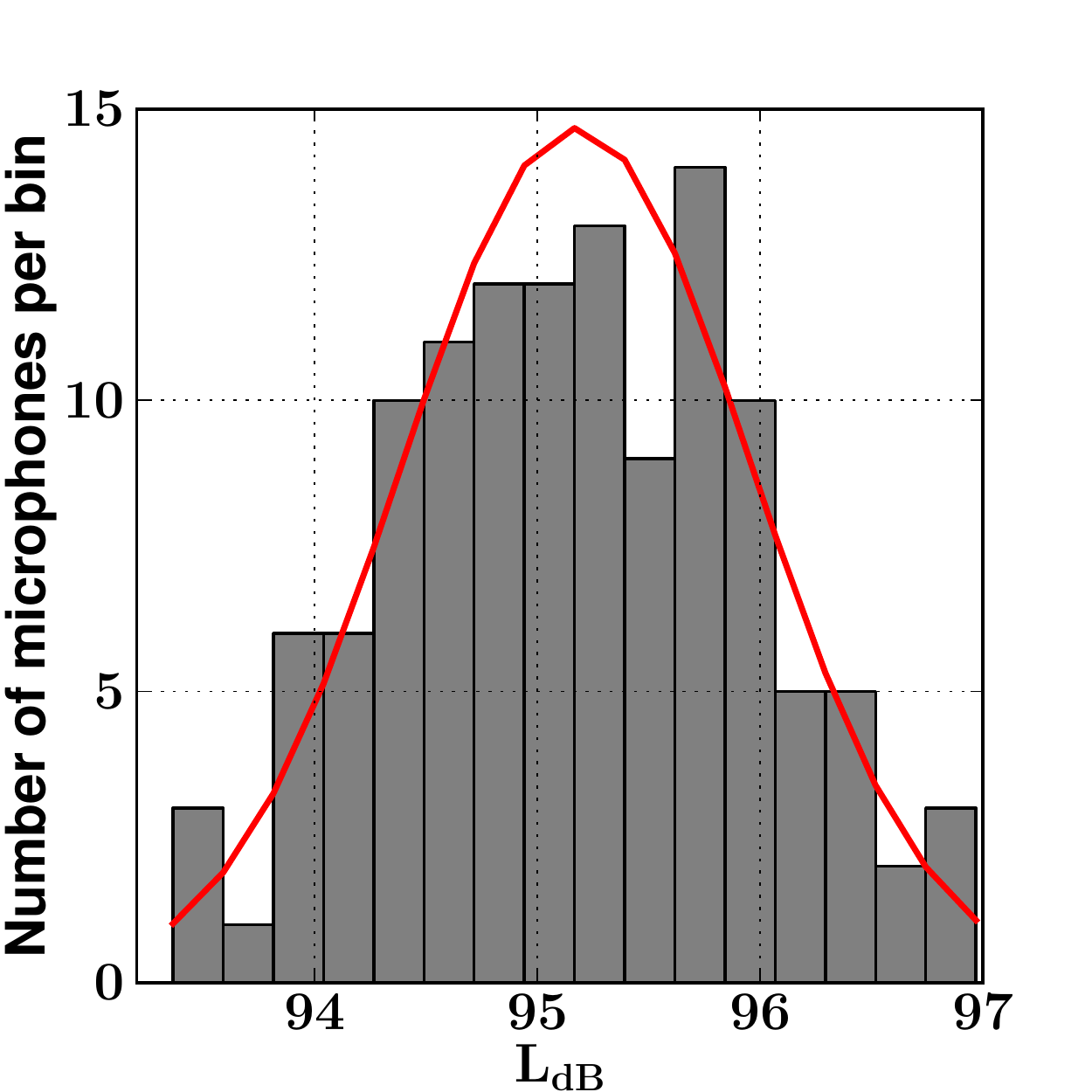}
	\label{fig:calib_sensitivity}
	}~~
	\subfloat[Normalized directivity in dB, individual patterns in grey, averaged pattern in red]{
	\includegraphics[width=0.4\textwidth]{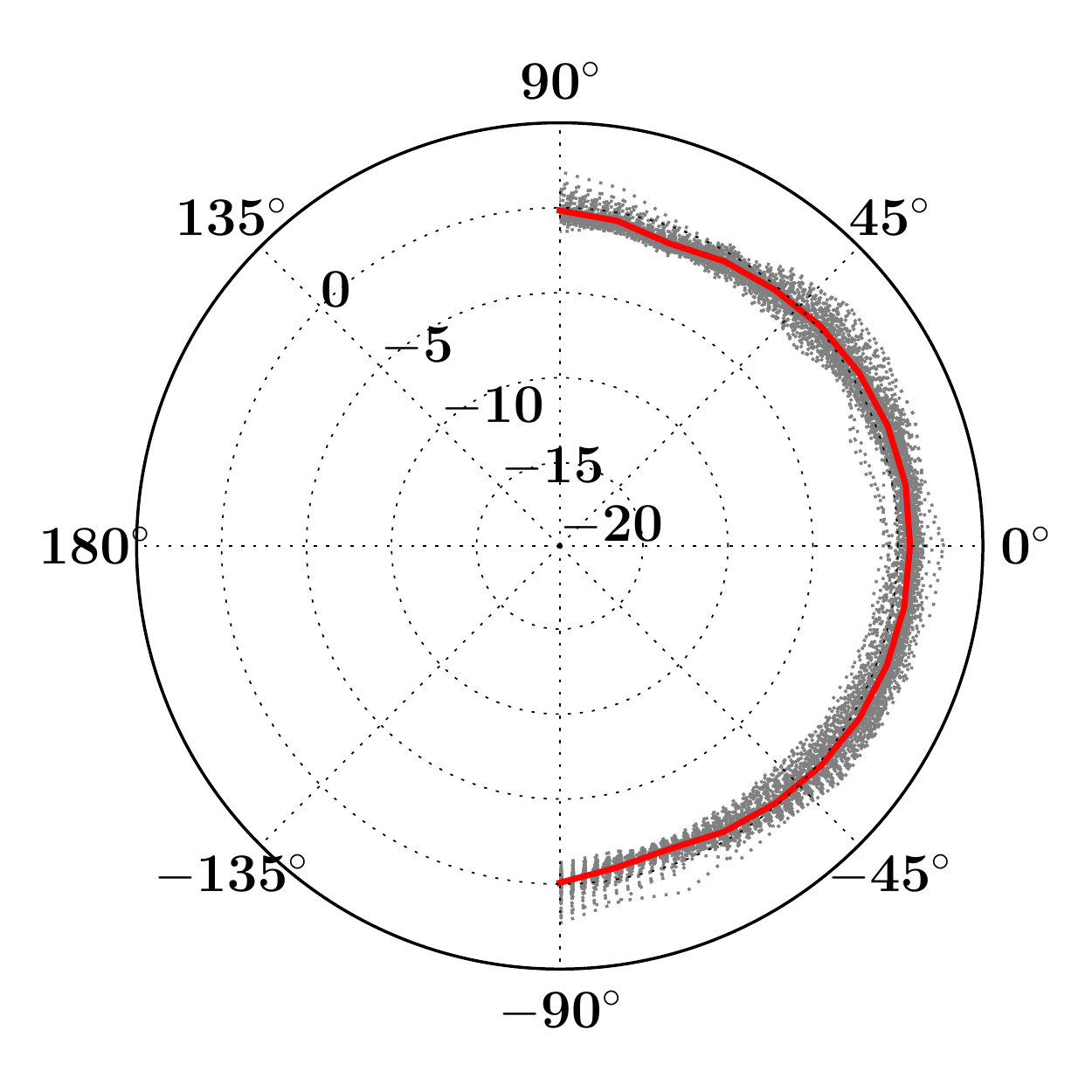}
	\label{fig:calib_directivity}	
	}
	\\
	\subfloat[Average narrow-band frequency response in dB referred to the reference microphone (red), and third-octave band levels (grey)]{
	\includegraphics[width=0.4\textwidth]{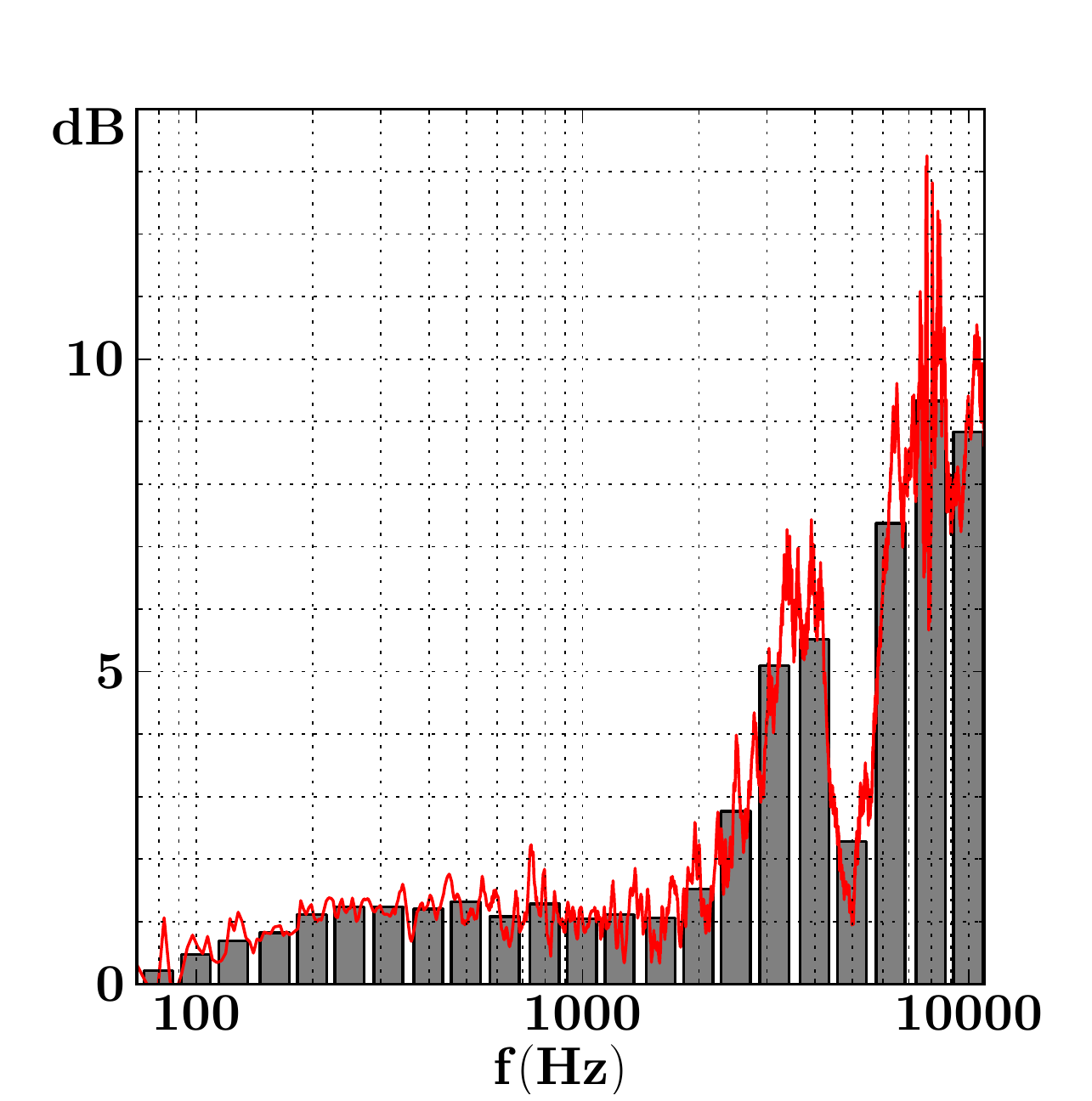}	
	\label{fig:calib_BP}
	}~~
	\subfloat[Self noise, individual spectra in dB referred to $2\cdot 10^{-5} \text{Pa/}\sqrt{\text{Hz}}$ (grey), averaged spectrum (red)]{
	\includegraphics[width=0.4\textwidth]{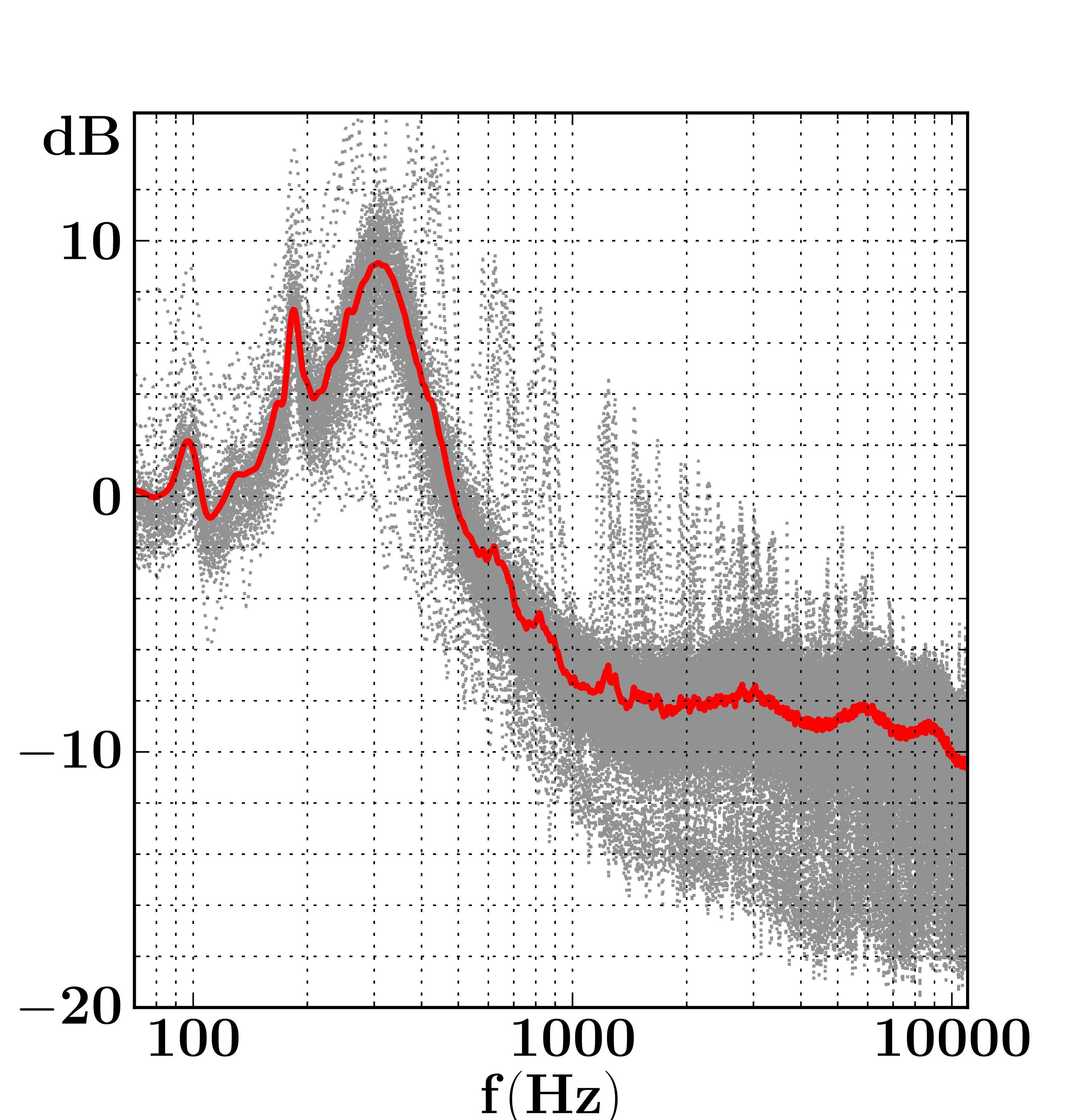}
	\label{fig:calib_SNR}
	}
\end{center}
\caption{Acoustic assessment of MEMS microphones: experimental results}
\label{fig:calib}
\end{figure}

In this section averaged results are presented in order to exhibit the global behaviour of MEMS microphones (fig. \ref{fig:calib}). Individual results are also presented in order to assess the statistical variation over the array.

The relative sensitivity variation is presented on the histogram, showing the 128 measured SPLs from the array during the experiment (fig. \ref{fig:calib_sensitivity}). This result straightly reflects the homogeneity among the microphones. In order to  quantify the variation, a normal distribution (red curve in the figure) is fit over the histogram, its standard deviation equals 0.8 dB. This small variation can easily be handled by means of simple equalization coefficients. 

The normalized frontal directivity is presented in figure \ref{fig:calib_directivity}; the average pattern shows a maximum $1.7$ dB deviation. Considering this small variation, the hypothesis of omni-directionality of the array is acceptable. Moreover the normalized individual patterns remain similar. So the averaged directivity pattern is representative of the global behavior of the microphones, and can be used if directivity correction is wanted according to the processed steering direction.

The frequency response is assessed on a frequency range which covers standard SLM octave bands implying typical community noises. Figure \ref{fig:calib_BP} shows the evaluated frequency response of MEMS microphones between $90$~Hz ($\approx 125/\sqrt{2}$~Hz) and $11$~kHz ($\approx \sqrt{2} \cdot 8$~kHz). In order to discard the loudspeaker frequency response, the transfer function of each MEMS microphone is normalized by the reference microphone response claimed to be flat into the studied frequency range. The variation remains lower than 2 dB up to 2 kHz, but beyond it becomes more significant, increasing by 7 dB up to 11~kHz. Note that frequency responses are homogeneous among the microphone set, so the variations should not affect imaging capabilities. However, it will bias the estimation of octave band SPLs; these discrepancies can be easily reduced in the frame of frequency-domain imaging, by equalizing the frequency response.

Listening to silence provides the self noise which is inherent to the acquisition system. It is directly obtained from the microphone output power spectra. The latters provide accurate information to estimate the SNR with respect to the frequency. As described in figure \ref{fig:calib_SNR} the resulting average spectrum pattern does not tally with a typical noise color. It is lower than $0 \text{ dB}$ (referred to $2\cdot 10^{-5} \text{Pa/}\sqrt{\text{Hz}}$) beyond $500$~Hz, and presents a maximum at $300$~Hz. Integrating this spectrum provides an overall 33.5 dB SPL acoustic noise, which is in accordance with the manufacturer specification (33 dB). Anyway, this specification does not lead to a global conclusion: it should be compared with the spectrum of the measured signal, so as to evaluate the SNR at each frequency bin.

Basically, the microphone set proves to have a good homogeneity. The 4 specifications meet the expectations and do not show any critical limit likely to affect the imaging process. Further results (in section \ref{sec:results}) are presented to validate experimentally the applicability of imaging with this array.

\section{Design of the realtime imaging processing scheme}
\begin{figure}
\centering
\includegraphics[width=0.30\paperwidth]{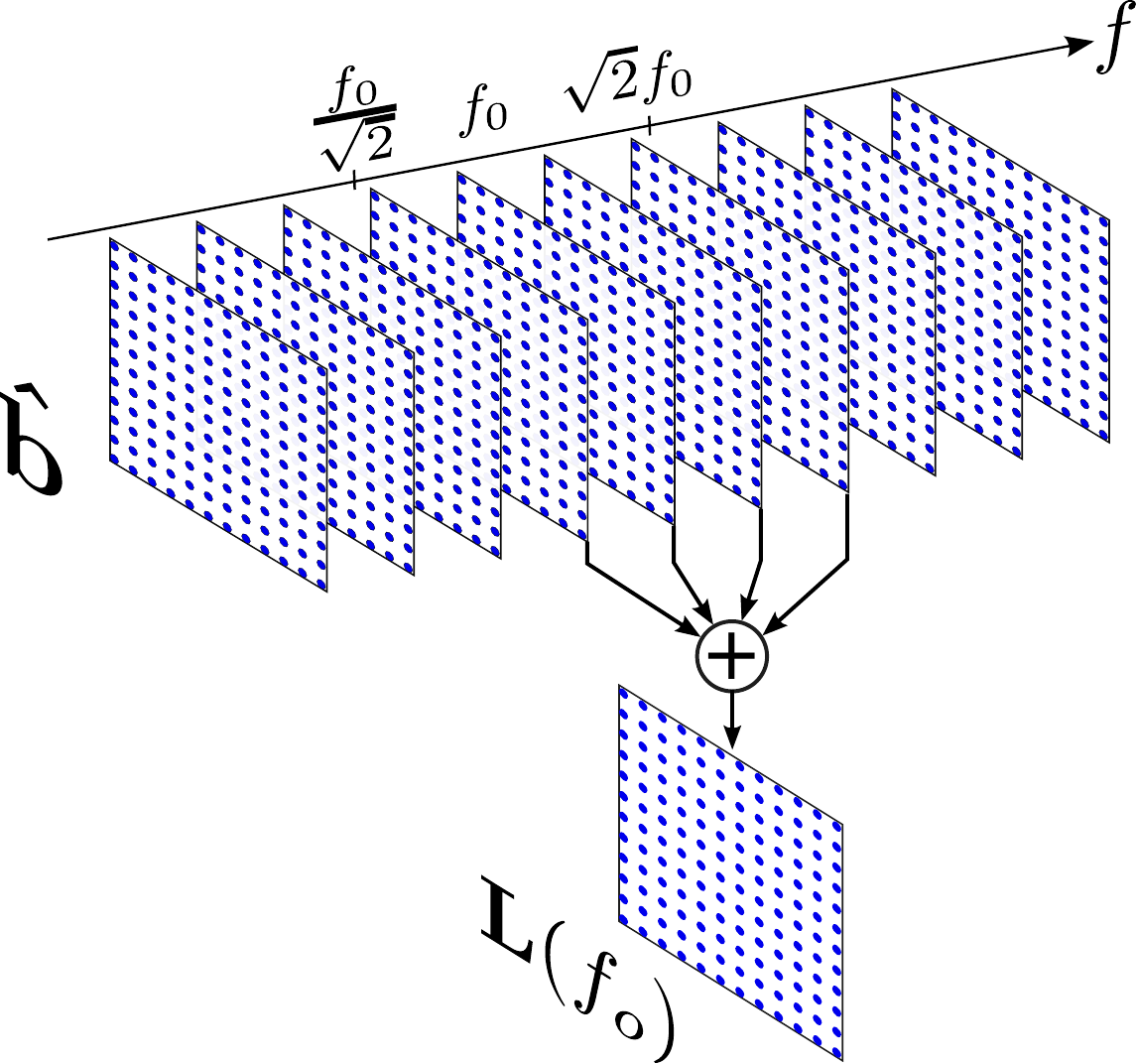}
\caption{Proposed imaging processing scheme}
\label{fig:WBBF}
\end{figure}

Imaging by octave bands needs successive stages (illustrated in fig. \ref{fig:WBBF}) of process of the basic data. The latter consist in $M$ microphone signals acquired continuously and resulting in the measured pressure matrix $\mathbf P$ to be processed:
\begin{equation}
\mathbf{P} = 
\left[ \begin{array}{c}
\mathbf{p}_1 \\
\vdots \\
\mathbf{p}_{M}
\end{array} \right]
=
\left[ \begin{array}{ccc}
p_{11} 		& \hdots & p_{1N_s}\\
\vdots 		&        & \vdots\\
p_{m1}& \hdots & p_{mN_s}\\
\vdots 		&        &\vdots\\
p_{M1}  	& \hdots & p_{MN_s}\\
\end{array} \right]  
\end{equation}

Each signal $\mathbf{p}_m$  contains the last $N_s$ pressure time samples. $\mathbf P$ is firstly beamformed in frequency domain to reconstruct the sound field at different steering points over the whole spectrum. Then, beamformed pressure is used to compute octave band SPLs of each steering point. 

\subsection{Broadband beamforming strategy}
This section exposes the adaptation of the BF method so as to provide images of the reconstructed sound field simultaneously, on the overall spectrum, according to a predefined grid of $N$ steering points.  

Let $\mathbf{x}_m$ and $\mathbf{y}_n$ be the positions of the $m$-th microphone and the $n$-th steering point respectively. We define $r_{nm}$, the component of the distance matrix $\mathbf R$:
\begin{equation}
r_{nm} = \Vert \mathbf{y}_n - \mathbf{x}_m \Vert 
\end{equation}

The broadband beamforming process goes through 2 steps.

First a Discrete Fourier Transform (DFT) is performed on the rows of $\mathbf{P}$ which gives $\mathbf{\hat P}$:  
\begin{equation}
\hat{\mathbf P} 
=
\left[ \begin{array}{c}
\hat{\mathbf p}_1 \\
\vdots\\
\hat{\mathbf p}_m \\
\vdots\\
\hat{\mathbf p}_M
\end{array} \right] 
=
\left[ \begin{array}{ccccc}
\hat{p}_{11} &\hdots & \hat{p}_{1k} &\hdots & \hat{p}_{1N_s}\\
\vdots & &\vdots &  &\vdots\\
\hat{p}_{m1} &\hdots & \hat{p}_{mk} &\hdots & \hat{p}_{mN_s}\\
\vdots & &\vdots &  &\vdots\\
\hat{p}_{M1} &\hdots & \hat{p}_{Mk} &\hdots & \hat{p}_{MN_s}\\
\end{array} \right]  
\end{equation}

The spectra $\hat{\mathbf p}_m$ have a frequency resolution $\Delta f = f_s / N_s$ and the frequency bins are located at $f_k = k \Delta f$.
For the frequency $f_k$ --- \textit{i.e.} from the $k$-th column of $\mathbf{\hat P} $ ---  the beamforming operation writes in matrix form: 
\begin{equation}
\mathbf{\hat b}_k
=
\mathbf{A}_k
\left[ \begin{array}{c}
\hat{p}_{1k}\\
\vdots \\
\hat{p}_{Mk}\\
\end{array} \right]  
=
\mathbf{A}_k
\mathbf{\hat{p}}_k\\
\label{eq:pim_mat}
\end{equation}
so from the array signal, the inverse wave propagator $\mathbf{A}_k$ gives the spatial distribution of sources at frequency $f_k$. Yet the emitted sources propagate up to the microphones by following the spatial wave equation. Assuming that radiated field is a distribution of monochromatic point sources, the solution of wave equation (\textit{i.e.} spherical wave) enables to build $\mathbf{A}_k$, whose elements are:
\begin{equation}
A_{k\;nm} = \dfrac{r_{nm}}{M} e^{\jmath 2\pi f_k\tfrac{r_{nm}}{c_o}  }
\end{equation}
$c_o$ being the speed of sound.

By extending the equation (\ref{eq:pim_mat}) over the whole spectrum, it finally provides the estimated source distribution $\mathbf{\hat b} = \left[\mathbf{\hat{b}}_{1} \ldots\mathbf{\hat{b}}_{k} \ldots \mathbf{\hat{b}}_{N_s} \right]$, at a $f_s / N_s$ frame rate.  

\subsection{Octave band imaging}
For the present development, SPL images according to standard octave bands were chosen so as to fit usual SLM data. The SPL of the grid in the octave-band $f_o$ results from summing beamformed data from $\mathbf{\hat b}$ along the frequency dimension included in the octave band (see fig. \ref{fig:WBBF}). Thus the SPL image at the $n$-th steering point is given by: 
\begin{equation}
\label{eq:Lim}
\mathbf{L}_{n} (f_o)= 10 \log \left( \dfrac{ 2\sum\limits_{f_o/\sqrt{2}}^{\sqrt{2}f_o} \Vert \hat{b}_{nk}\Vert^2}{N_s^2 p_o^2}  \right)
\end{equation}
with $p_o=2.10^{-5}$~Pa in air. At last, the stantard A-, B- or C- weighting can be applied. Otherwise data is directly displayed on a chosen dynamic range.

Although the previous process concerns an octave-band visualisation, it can be straightforwardly extended to a third-octave band, or any relevant set of frequencies  according to the wanted diagnosis. 

\subsection{GPU implementation}
\label{sec:GPUImplt}

Realtime beamforming is achievable on CPU at a single frequency: if imaging is performed for a grid of $N$ steering points, then it involves computing $N$ elements of $\mathbf{\hat b}$ \textit{i.e.} one $\mathbf{\hat b}_k$. However, extending the same process onto multiple frequencies strongly increase the necessary computations, and CPU architecture is not adequate for this purpose.


Yet one should notice that the computing steps providing the matrix  $\mathbf{\hat b}$ are the same whatever the considered steering point and frequency bin (\textit{i.e.} the row and the column of $\mathbf{\hat b}$). Only the input data $\mathbf{P}$ vary. This framework calls for the use of parallel computing and leads in particular to the implementation of  the SIMD (Single Instruction Multiple Data) strategy.

Several parallel computing hardwares exist, but we favor low global cost, the possibility of an embedded use, and an efficient development stage according to \citep{Chen2012}. Nowadays GPUs become attractive in order to reach such capabilities. Recent coming of  development tools offers the possibility to extend fully custom parallel programming to classic graphical features, and lead to a new programming branch: General Purpose processing on GPU \cite{Kirk2010}. Besides, usual handling of graphics for image display using textures remains as easy and efficient, \textit{e.g.} by using \textit{OpenGL} libraries \cite{Nvidia2013}.

GPU is an independent system plugged to the PCIe bus of the host computer. It includes its own memory where the measured data is to be prior transferred. Besides the process must be adapted to the GPU parallel architecture. Nvidia's \textit{CUDA} platform has been chosen, even though \textit{OpenCL} standard theoretically permits the same approach on different kinds of GPU architectures.

Realtime capability is achieved by displaying images continuously to the screen. The operation producing each octave-band SPL image goes through 5 steps:

\begin{enumerate}[Step 1.]
\item after $N_s / f_s$ seconds, the signals $\mathbf{P}$ are transferred from the host memory to the GPU's global memory. \label{itemGPU1}
\item $\mathbf{P}$ undergoes the DFT giving $\mathbf{\hat P}$, by performing $M$ parallel FFTs (one for each row of $\mathbf{P}$). \label{itemGPU2} 
\item The SIMD strategy affects one thread to the computation of one $b_{nk}$ and finally yields the required $\hat{\mathbf{b}}$. \label{itemGPU3}
\item The SIMD strategy is still opted to obtain the octave-band SPLs: one thread independently computes Eq \ref{eq:Lim}, \textit{i.e.} one element of $ \mathbf{L}(f_o)$. \label{itemGPU4}
\item Octave-band pressure levels are cast into a RGBA texture array, thanks to the \textit{CUDA} compatibility with the \textit{OpenGL} library. This texture corresponds to the final displayed image.
\end{enumerate}

The complete program consists of a \textit{python} script (OS-independent), with user configurable initial parameters (problem geometry, and frequency bands to be displayed) and  realtime tunable parameters (dynamic range and maximum SPL).

For realtime performances, the host CPU is not involved in computing. Its role is limited to handling tasks regarding USB communication, and GPU control. Communication with the acquisition system is achieved by means of the \textit{libusb} library. The chosen USB protocol allows asynchronous transfer (\textit{i.e.} data reception is performed in non-blocking background processes) so that CPU remains free to both execute GPU controls and meet the transfer rate requirements of USB2.0 bus. Afterwards, CPU launches the successive GPU global processing kernels (steps \ref{itemGPU2}, \ref{itemGPU3} and \ref{itemGPU4}) with \textit{pycuda} \citep{kloeckner_pycuda_2012}. The final texture display uses the \textit{FreeGLUT} library, which affords a simplified use of \textit{OpenGL}, and also allows display and keyboard events control for realtime tuning.


\subsection{Meeting the realtime requirements}

The benchmark of the parallel process is led for two different GPUs:
\begin{inparaenum}[(i)]
\item a Nvidia Titan (with 2688 CUDA cores) and 
\item a Nvidia Quadro 2000 (192 cores).
\end{inparaenum}
The similar beamformer has been adapted to sequential execution, and implemented for CPU (Intel Xeon X5570 processor) to assess the improvement brought by GPU. With a view to comparing consistent execution times, it should be noted that CPU beamformer is compiled from a C program, and executed on a single core. 

Realtime capability is evaluated for a specific configuration: computing one octave band image of $N=10^4$ pixels, at $f_o = 1$ kHz, from a $125$ ms acquisition of the 128  microphone signals sampled  at $f_s = 50$ kHz. Since the data flow is received continuously from the acquisition system, reaching realtime is possible if the computation needs less than $125$ ms to process the whole data. Table \ref{tab:benchmark} shows processing times, all tasks included (steps \ref{sec:GPUImplt}.\ref{itemGPU1} to \ref{sec:GPUImplt}.\ref{itemGPU4}),
GPU implementation presents a significant increase compared with CPU implementation. In the given conditions of the benchmarked example, realtime is possible for both GPU, but not for CPU.

Therefore realtime is reachable, but computation time changes if the configuration changes. Its complexity is experimentally confirmed to be in coherence with theory, and two parameters should especially be taken into account for realtime performance: the number of steering points $N$ (proportionally increasing the computation time), and the frequency center of the octave band. Doubling $f_o$ doubles time computation then the highest time consuming imaging happens in high frequency. This is due to the amount of frequency bins in an octave band, geometrically increasing as a function of $f_o$.

\begin{table}
\centering
\begin{small}
\begin{tabular}{|c||c|c|c|}
\hline 
 & CPU  & GPU Quadro 2000 & GPU Titan \\ 
\hline 
Processing time & 8.12 s & 31.1 ms  & 4.96 ms \\ 
\hline 
Computing capability (pix/s) & 1231 & 321543 & 2016129 \\ 
\hline 
GPU-CPU speed gain  & / & 261 & 1637 \\ 
\hline 
\end{tabular} 
\caption{Benchmark results:  Computation of 1 octave-band image of ($N = 10^4$, $f_o = 1$ kHz, $f_s = 50$ kHz, $M = 128$, $N_s = 6144$.}
\label{tab:benchmark}
\end{small}
\end{table}

\section{Imaging results}
\label{sec:results}

Through two scenarios, the diagnosis abilities of the SLM-like camera are investigated and illustrated:
\begin{itemize}
\item the first scenario is de facto a calibration experiment. The proper functioning of the camera and absolute estimation of SPLs is gauged. To this purpose, a simple static configuration is set, with two speakers emitting monochromatic sources of known frequencies and power. Their SPL is separately measured with a SLM (\textit{Br\"{u}el \& Kjaer 2236} of type I) and compared with the values estimated on images.
\item The second scenario illustrates the system by diagnosing a vacuum cleaner. As expected, it is able to discriminate different sources of noise (motor, intake port, exhaust). From a global analysis of the displayed images, the interest of the proposed system of diagnosis is discussed.
\end{itemize}

The final camera software tends towards the classic SLM properties. Indeed SLM standards are followed: octave bands are centered at standard frequencies according to the \textit{ISO 266} norm, and the images allowing diagnosis on a 6 octave basis are refreshed simultanously at a 125 ms rate. The system thus provides an effective broadband and spatial diagnosis capability in realtime. 

\begin{figure}
\begin{center}
\includegraphics[width=0.45\textwidth]{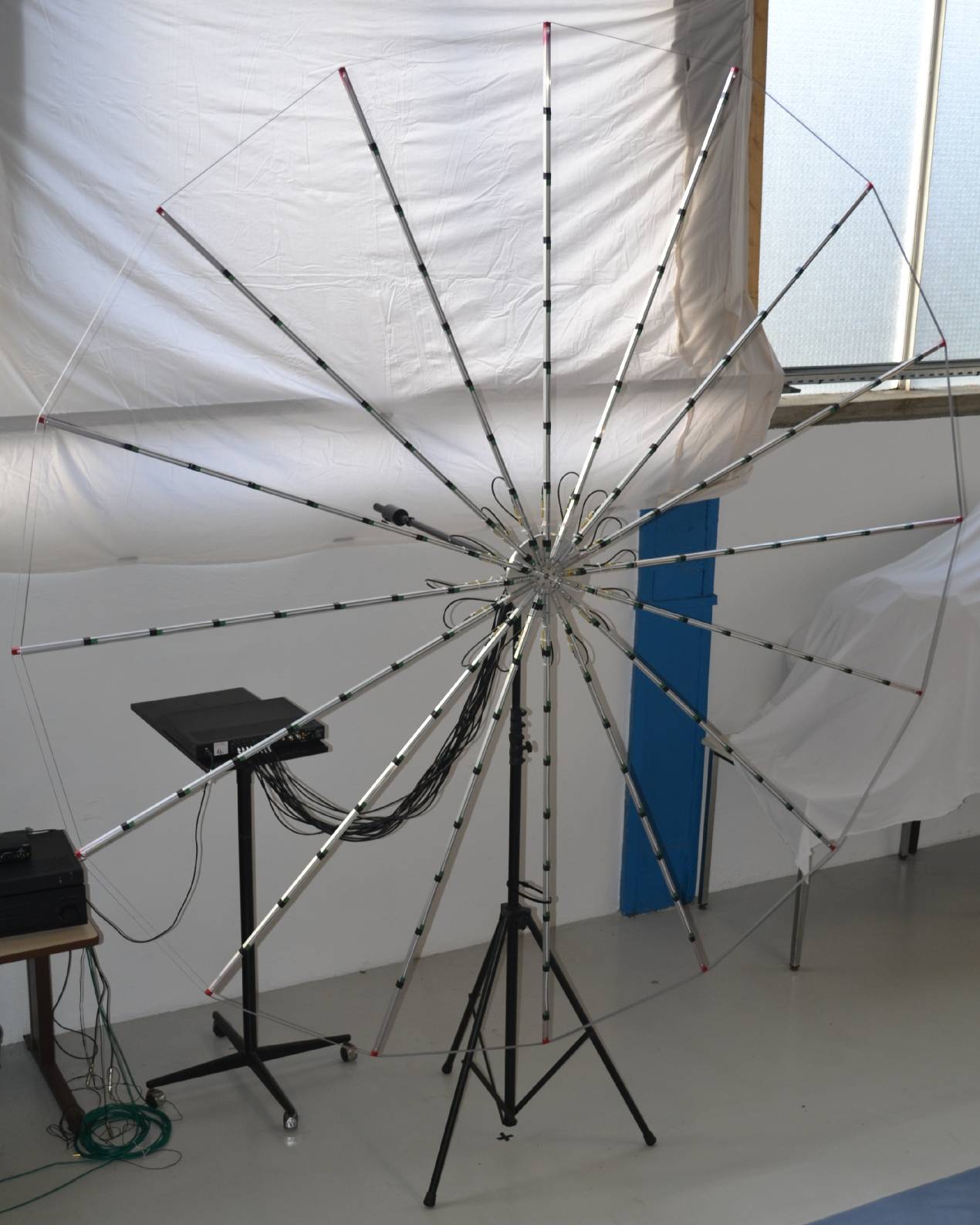}
\end{center}
\caption{The data acquisition system configuration during experiments}
\label{fig:antenna}
\end{figure}

The acquisition system configuration is fixed for all experiments: the array of 128 MEMS microphones is a circular antenna (figure \ref{fig:antenna}), and sampling pressure is set at $f_s=50$~kHz. The software configuration keeps a steering grid made of $N = 100\times100$ points, however the parameters defining the grid geometry are adapted to each experiment. At last the host computer is the same as the one used for benchmark, and Nvidia GTX Titan GPU is chosen to guarantee better performances.

\subsection{Calibration of the camera for absolute level measurement}

The complete calibration implies tests over the whole spectrum, and induces a great quantity of results. The exposed calibration experiments are limited to a specific disposition. Two identical speakers are located 5 meters away from the circular array. Both emit a source of same nature: a $3$~kHz sinus at constant power. Three different tests are performed, where:
\begin{itemize}
\item the right speaker always emits at a SPL of $69$~dB
\item the left speaker respectively emits at $69$, $64$ and $59$~dB. 
\end{itemize}
In this way, the absolute level of each source is gauged, as well as the relative level between the two speakers in the same frame. Assuming a spherical wave propagation, the same levels are expected to be found in the octave-band images as well. In order to confirm the right emission level, each emitting source is measured separately with the SLM. The latter is placed 1 meter in front of the source, and calculates the Equivalent Continuous Sound Level ($L_{\text{eq}}$) over 30 seconds.

\begin{figure}
\begin{center}
	\subfloat[SPL of loudspeakers: $69$~dB (left and right)]{
	\includegraphics[width=0.3\textwidth]{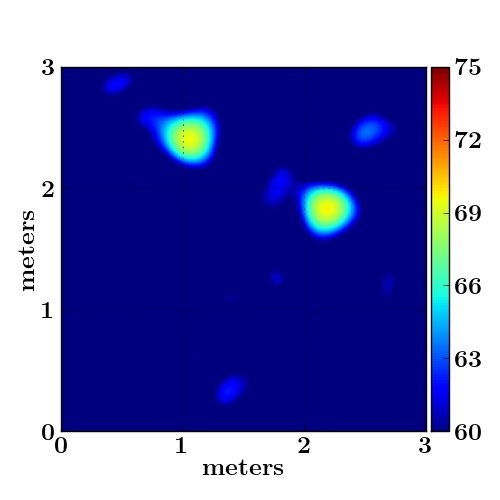}
	\label{fig:calibCam_1}
	}~~
	\subfloat[SPL of loudspeakers: $64$~dB (left), $69$~dB (right)]{
	\includegraphics[width=0.3\textwidth]{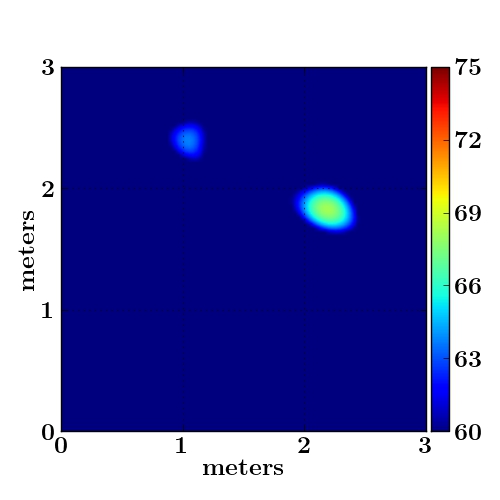}
	\label{fig:calibCam_2}	
	}~~
	\subfloat[SPL of loudspeakers: $59$~dB (left), $69$~dB (right)]{
	\includegraphics[width=0.3\textwidth]{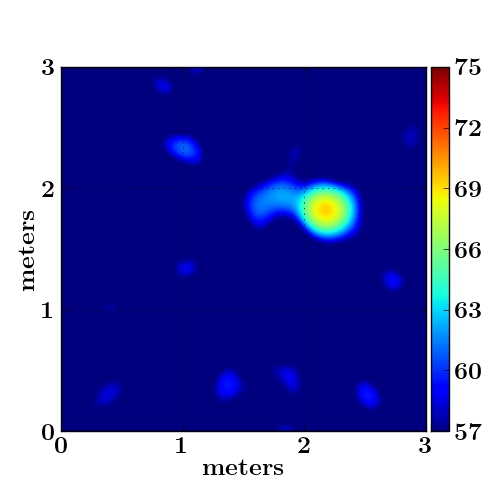}	
	\label{fig:calibCam_3}
	}
\end{center}
\caption{Calibration of the camera, with two speakers emitting a $3$~kHz sine}
\label{fig:calibCam}
\end{figure}

Figure \ref{fig:calibCam} presents the $4$~kHz octave band image, for the three different tests. Table \ref{tab:SPLs} summarizes the results, and compares the assessed SPLs with the expected ones. The absolute SPLs in images deviate by less than $0.8$~dB except in one case: the left speaker in (c), which deviates by $1.9$~dB. Note that this one is the lowest SPL emitted among the tests. Moreover figure \ref{fig:calibCam_3} shows some lobes appearing with a similar acoustic level. These lobes are inherent to the array geometry and the BF. They exhibit the intrinsic dynamic range of the used antenna, which is therefore unable to discriminate sources whose SPLs differ by more than $12$-$13$ dB. Nevertheless this dynamic range is characteristic of the array and the frequency at which imaging is performed. It may be improved by designing another antenna geometry. As a result, the absolute levels provided by the camera remain of good accuracy. But it starts to degrade for low level sources, if their relative level with the highest source becomes too significant.

\begin{table}
\begin{center}
\begin{tabular}{|c||c|c|c|}
\hline 
Test & (a) & (b) & (c) \\ 
\hline 
Left speaker level & $69.6$ ($69.0$) & $63.8$ ($64.0$) & $60.9$ ($59.0$) \\ 
\hline 
Right speaker level & $69.8$ ($69.0$) & $68.2$ ($69.0$) & $69.3$ ($69.0$) \\ 
\hline 
Relative level & $0.2$ ($0$) & $4.5$ ($5.0$) & $8.4$ ($10.0$) \\ 
\hline 
\end{tabular} 
\end{center}
\caption{Calibration of the camera. Measured acoustic levels of sources in dB, presented in the following format: Camera SPL (SLM $L_{\text{eq}}$)}
\label{tab:SPLs}
\end{table}

\subsection{Illustration experiment: diagnosing a vacuum cleaner}

\begin{figure}[t]
\centering
\begin{center}
\includegraphics[width = 0.99 \linewidth]{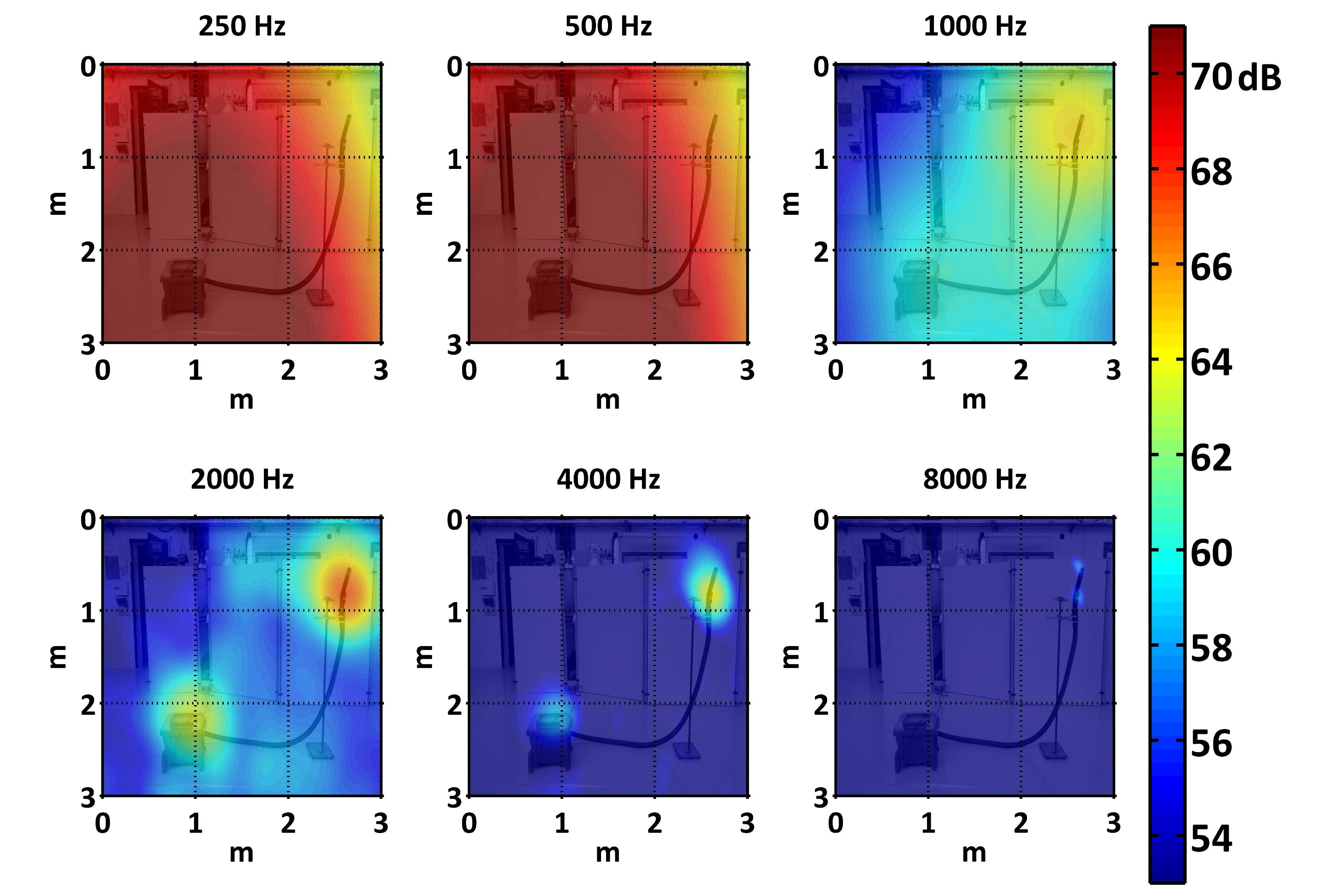} 
\end{center}
\caption{Diagnosis experiment: imaging of the vacuum cleaner at 6 octave bands. SPLs are referred to $5 \cdot 10^{-5}$ Pa.}
\label{fig:Vacuum}
\end{figure}

The second scenario intends to diagnose the noise from a working vacuum cleaner. It is located 4 meters away from the array. The realtime images of the camera shown in figure \ref{fig:Vacuum} map the radiated field in a $3 \text{~m} \times 3 \text{~m}$ grid.

A first global overview validates three expected aspects:
\begin{itemize}
\item As in the case of a monochromatic acoustic camera, imaging the radiated sound enables the localization of different sources in space. Here they originate essentially from two parts: the motor and the inlet of the hose.
\item Octave band Imaging enables a finer analysis, and leads to an accurate definition of the sources over the spectrum. Such information orientates the diagnosis to a possible identification of the source nature. For instance, low frequency noise indicates the noise due to the motor rotation. At middle and high frequencies, aeroacoustic noise from the intake port emits on a broader band. Besides, note another significant source in the same frequency range, from the motor part. One possible interpretation is that it is aeroacoustic noise from the exhaust port.
\item absolute octave band SPLs can be estimated for each source of noise. In addition to the classic functionality of a SLM, the camera is able to provide the level of each source separately. Each pixel of the image can be interpreted as a virtual octave-band SLM located at each steering point of the grid. In the present case, the noisiest source is from the motor at low frequency, however images indicate that the intake port prevails in middle and high frequencies.
\end{itemize}
However, the main disadvantage of BF clearly emerges: the beamwidth varies according to frequency. It results in a low resolution in images centred at $250$ and $500$~Hz: the motor noise appears by a large spot. Conversely, high frequency images provide a good resolution, and a fortiori improve the resolving capacity. The 8 kHz octave band is a good example, where two close sources are separated: the main intake port and the suction adjuster valve.

This experiment illustrates the capability of the imaging system to merge the functionality of a SLM and acoustic imaging. Furthermore, realtime capability brings the advantage of directly interacting with the experiment. Anyway, the present diagnosis provides a rich and concise information, which allow the analysis to be fast and convenient.

\section{Conclusion}

We presented a complete diagnosis tool merging the capabilities given by imaging and SLM diagnosis. The solution includes both the hardware and the software requirements, from pressure measurements to the display of the processed data. The main purpose was to propose an acoustic camera which affords to give an intelligible description of the radiated sound. We chose to follow the same standards as the SLM, even though other configurations are customizable if it is favourable to the experiment.

The priorities were to have a convenient hardware system, and to work in realtime. We benefited from technologies which overcome the current limits of acoustic imaging systems. Digital MEMS microphones allowed us to build an acquisition system, whose global architecture is less complex and more flexible than conventional imaging ones. The characterization experiments confirmed that these components are attractive for aerial acoustic imaging of community noise. They have a good homogeneity and, if necessary, the existing discrepancies such as the frequency response are easily balanced.

The use of GPU proved to be adapted to our application. Its strong capacity in parallel computing fits with the whole implemented process. The high performances enable to achieve the execution of the program in realtime, which is an important criterion regarding diagnosis possibilities. Additionally, a GPU remains compact and then favors the possibility of having an embedded system.

The experiments showed that such a tool opens up to new perspectives of diagnosis. Indeed they proved that SLM feature and acoustic imaging complete each other, although both tools are initially distinct and used for different purposes. Yet, even though Beamforming was chosen, other algorithms exist, improving image quality \textit{e.g.} resolution. Despite the need for more computation, such algorithms could be interesting alternatives according to the concerned experiment.

\section{Acknowlegement}

The authors thank the LNE for making their facilities available, where microphone characterization experiments have been performed.


\bibliographystyle{elsarticle-num}
\bibliography{jabref_database}




\end{document}